# Enhancement of thermoelectric properties by Se substitution in layered bismuth-chalcogenide LaOBiS$_{2-x}$Se$_x$

Yoshikazu Mizuguchi[1]*, Atsushi Omachi[1], Yosuke Goto[2], Yoichi Kamihara[2], Masanori Matoba[2], Takafumi Hiroi[1], Joe Kajitani[1], Osuke Miura[1]

1. *Department of Electrical and Electronic Engineering, Tokyo Metropolitan University, 1-1, Minami-osawa, Hachioji 192-0397, Japan*
2. *Department of Applied Physics and Physico-Informatics, Keio University, 3-14-1 Hiyoshi, Yokohama 223-8522, Japan*

Corresponding author: Yoshikazu Mizuguchi (mizugu@tmu.ac.jp)

*Abstract*

We have investigated the thermoelectric properties of the novel layered bismuth chalcogenides LaOBiS$_{2-x}$Se$_x$. The partial substitution of S by Se produced the enhancement of electrical conductivity (metallic characteristics) in LaOBiS$_{2-x}$Se$_x$. The power factor largely increased with increasing Se concentration. The highest power factor was 4.5 μW/cmK$^2$ at around 470 °C for LaOBiS$_{1.2}$Se$_{0.8}$. The obtained dimensionless figure-of-merit (*ZT*) was 0.17 at around 470 °C in LaOBiS$_{1.2}$Se$_{0.8}$.

Discovery of novel thermoelectric materials with a high performance is one of the most important issues for the development of thermoelectric application [1]. In recent years, researchers have focused on layered materials as candidates of a novel thermoelectric material because their low-dimensional crystal structure and electronic states can produce a high thermoelectric property as realized in $Bi_2Te_3$, cobalt oxides and $CsBi_4Te_6$ [2-4].

Since 2012, layered compounds with the $BiS_2$ layers have got much attention in the field of condensed matter physics due to the discovery of superconductivity with a transition temperature as high as 11 K in $Bi_4O_4S_3$ [5], $RE$O$_{1-x}$F$_x$BiS$_2$ ($RE$ = La, Ce, Pr, Nd and Yb) [6-12] and $AE_{1-x}RE_x$FBiS$_2$ ($AE$ = Sr and Eu) [13-16]. Their crystal structure is basically composed of alternate stacks of the conduction layer ($BiS_2$ layer) and the blocking layer such as REO layer. In the $BiS_2$-based materials, the Bi-$6p$ orbitals within the $BiS_2$ layer are essential for electronic conduction. It has been revealed that the electronic states are easily tunable by element substitutions at the blocking layers. Having considered the low-dimensional crystal structure and electronic states, we have regarded the layered bismuth chalcogenides as potential thermoelectric materials.

In the previous study, we investigated physical properties of the $BiS_2$-based layered compounds LaO$_{1-x}$F$_x$BiS$_2$ at high temperatures [17]. We observed an anomalous metallic behavior in LaOBiS$_2$: a metal-semiconductor transition was revealed at around 270 K. The highest power factor of 1.9 μW/cmK$^2$ at ~480 °C was observed for LaOBiS$_2$. In addition, we investigated the effect of F substitution (electron doping) on the power factor in LaO$_{1-x}$F$_x$BiS$_2$. Although the electrical conductivity increases by the partial substitution of O by F, the Seebeck coefficient largely decreases. Hence, the F substitution resulted in a decrease of power factor. This result suggests that the element substitution at the blocking layer negatively works in enhancing thermoelectric properties. Therefore, we have investigated the effects of element substitution at the $BiS_2$ conduction layer on thermoelectric properties of LaOBiS$_2$. Here, we report a large enhancement of the power factor and the dimensionless figure-of-merit ($ZT$) by a partial substitution of S by Se in LaOBiS$_{2-x}$Se$_x$. On the basis of the systematic changes in lattice constants, we have considered that the S and Se ions are mixed as shown in Fig. 1(a). The partial Se substitution achieves both metallic conductivity and a large Seebeck coefficient in LaOBiS$_{2-x}$Se$_x$.

Polycrystalline samples of LaOBiS$_{2-x}$Se$_x$ were prepared by the solid-state reaction method using powders of La$_2$S$_3$ (99.9 %), La$_2$O$_3$ (99.9 %), Bi$_2$S$_3$, Bi$_2$Se$_3$ and grains of Bi (99.999 %). The Bi$_2$S$_3$ and Bi$_2$Se$_3$ powders were synthesized by reacting Bi, S (99.99 %) or Se (99.99 %) grains in an evacuated quartz tube. Other chemicals were purchased from Kojundo-Kagaku Laboratory. The mixture of starting materials with nominal compositions of LaOBiS$_{2-x}$Se$_x$ ($x$ = 0, 0.2, 0.6, 0.8 and 1.0) was mixed-well, pelletized and sealed into an evacuated quartz tube. The samples were heated at 700 ºC for 15 h. The obtained products were ground, pelletized, sealed into an evacuated quartz tube and heated under the same heating conditions for homogenization. The prepared samples were characterized by powder x-ray diffraction (XRD) with a CuKα radiation using the $\theta$-2$\theta$ method.

The XRD patterns of the typical LaOBiS$_{2-x}$Se$_x$ samples ($x$ = 0 and 1.0) are shown in Fig. 1(b). Almost single-phase samples were obtained for $x$ = 0 - 1.0, except for a small amount of La-oxide impurities as indicated by asterisk in the XRD pattern. The numbers in the XRD pattern is Miller indices with the space group of *P*4/*nmm*. Although we attempted to prepare the samples with $x$ = 1.5 and 2.0, single-phase samples were not obtained probably due to the existence of a solubility limit of Se for the S site in LaOBiS$_{2-x}$Se$_x$. The lattice constants of $a$ and $c$ were calculated using the peak positions of the (200) and (003) peaks, and plotted in Fig. 1(c) and 1(d), respectively. We have corrected the peak positions using the corrected data obtained from the peak position of Si [14]. Both the lattice constants of $a$ and $c$ increase with increasing Se concentration. The increase of lattice volume can be understood by the difference in the ionic radius of S$^{2-}$ and Se$^{2-}$.

The electrical resistivity and the Seebeck coefficient were measured by the four-terminal method using ULVAC-RIKO ZEM-3 up to 480 ºC in an atmosphere of low-pressure He gas. The measurements of the thermal conductivity ($\kappa$) was conducted at temperatures up to 400 ºC using a lamp heating unit (Ulvac, MILA-5000). The $\kappa$ was obtained from the slopes of the plots of heat flux density vs. $\Delta T$ using a strain gauge as a heater and was measured under pressures less than 10$^{-3}$ Pa. The heat loss by radiation through the samples [18] was subtracted under the assumption that emissivity is independent of temperature and wavelength during the measurements of $\kappa$. The emissivity of 0.8 was employed on the basis of reflectivity measurements, which were performed at room temperature using a spectrometer equipped with an integrating sphere (Hitachi High-tech, U-4100). The relative density of the sintered pellet used in these measurements is ~85 % (84 % for $x$ = 0 and 91 % for $x$ = 1.0).

The power factor was calculated using an equation $P = S^2/\rho$, where $S$ and $\rho$ are Seebeck coefficient and electrical resistivity, respectively. The dimensionless figure-of-merit was calculated using an equation of $ZT = P\,T\,/\,\kappa$, where $\kappa$ and $T$ are thermal conductivity and absolute temperature, respectively.

Figure 2(a) shows the temperature dependences of electrical resistivity for LaOBiS$_{2-x}$Se$_x$. An anomaly is observed at around 400 ºC for $x$ = 0.2, which is similar to those observed at around 200 ºC in $x$ = 0 (LaOBiS$_2$) and F-doped LaOBiS$_2$ [17]. The anomaly may be related to a charge-density-wave transition, which was recently observed in EuFBiS$_2$ [16]. In the temperature dependences for $x$ = 0.6, 0.8 and 1.0, metallic behavior is observed without any anomalies below 480 ºC. With increasing Se concentration, the values of electrical resistivity systematically decrease, indicating that the Se substitution enhances metallic characteristics. The value of electrical resistivity at around 470 ºC for LaOBiSSe ($x$ = 1.0) is ~4.7 mΩcm, which is clearly lower than that for LaOBiS$_2$ ($x$ = 0) (~21 mΩcm) [17]. To clarify the mechanism of the enhancement of electrical conductivity by Se substitution, investigations with single crystals and band calculations are needed.

Figure 2(b) shows the temperature dependences of Seebeck coefficient for LaOBiS$_{2-x}$Se$_x$. All the samples show negative Seebeck coefficient, indicating that the contributing carrier is basically electron. The absolute value of the Seebeck coefficient tends to increase with increasing temperature for all the samples. Although the absolute values of Seebeck coefficient decrease with increasing Se concentration, the values at around 470 ºC for $x$ = 0.8 and 1.0 are still around -150 μV/K, which is a relatively high value as a metallic compound.

The obtained power factors are plotted in Fig. 2(c) as a function of temperature. The values of power factor are largely enhanced by Se substitution. For $x$ = 0.2 and 0.6, the temperature dependence of power factor shows a maximum at around 200 - 300 ºC. For $x$ = 0.8 and 1.0, the values of power factor monotonously increase with increasing temperature within a range of $T$ < 480 ºC. The highest power factor is 4.5 μW/cmK$^2$ at around 470 ºC for $x$ = 1.0 (LaOBiSSe). This value is 237 % of the highest value obtained in LaOBiS$_2$.

Finally, we discuss the thermoelectric performance of the LaOBiS$_{2-x}$Se$_x$ system. In order to calculate the dimensionless figure-of-merit ($ZT$), we have measured thermal conductivity for the end members ($x$ = 0 and 1.0). Figure 3(a) shows the temperature dependences of thermal conductivity for $x$ = 0 and 1.0. It is found that the thermal conductivity for LaOBiS$_{2-x}$Se$_x$ is almost independent of temperature. Furthermore, the obtained values of thermal conductivity are almost the same for LaOBiS$_2$ and LaOBiSSe, indicating that the partial substitution of S by Se does not affect the thermal conductivity in this system. In general, electronic thermal conductivity increases with enhancing metallic characteristics. On another front, the phonon thermal conductivity could decrease by substituting the chalcogen site in LaOBiS$_{2-x}$Se$_x$ due to introduction of the randomness. Hence, it is considered that the total thermal conductivity does not change with increasing Se concentration because of the compensation of the changes in the electron thermal conductivity and the phonon thermal conductivity. Therefore, we use a typical (average) value of $\kappa = 2$ W/m·K, which is the value estimated at room temperature for LaOBiS$_2$ and LaOBiSSe, to estimate $ZT$. This $\kappa$ value is consistent with that reported in the previous study on low-temperature thermal properties of LaOBiS$_2$ [19]. The temperature dependences of dimensionless figure-of-merit ($ZT$) for LaOBiS$_{2-x}$Se$_x$ are plotted in Fig. 3(b). For all the samples, the value of $ZT$ tends to increase with increasing temperature, and it reaches a maximum value at around 470 ºC, which is the measurement limit of the present work. This fact indicates that the LaOBiS$_{2-x}$Se$_x$ system is a potential thermoelectric material for the use at high temperatures. The highest value of $ZT$ is 0.17 for $x = 0.8$ (LaOBiS$_{1.2}$Se$_{0.8}$) at 470 ºC. To further increase the thermoelectric performance in this family, precise tuning of carrier concentration and/or local crystal structure are needed. In this family, partial substitutions at the blocking layer can tune both the carrier concentration within the conduction layers and the local structure that largely affects the physical properties [6, 17, 20]. Therefore, a strategy to enhance the thermoelectric performance of this system will be tuning of the structure of the blocking layer.

In conclusion, we have synthesized polycrystalline samples of novel layered bismuth chalcogenides LaOBiS$_{2-x}$Se$_x$ and systematically investigated thermoelectric properties. It was found that a partial substitution of S by Se enhanced metallic conductivity. The power factor largely increased with increasing Se concentration. The highest power factor was 4.5 μW/cmK$^2$ at around 470 ºC for LaOBiS$_{1.2}$Se$_{0.8}$. We found that the thermal conductivity for LaOBiS$_{2-x}$Se$_x$ is independent of both temperature and Se concentration. Using an average value of thermal conductivity, $\kappa = 2$ W/m·K, we calculated the dimensionless figure-of-merit ($ZT$) as a function of temperature. The highest $ZT$ was 0.17 at around 470 ºC in LaOBiS$_{1.2}$Se$_{0.8}$. Optimization of the carrier concentration and/or the local structure will further enhance the thermoelectric performance of the layered bismuth chalcogenides.

*Acknowledgements*

This work was partly supported by a Grant-in-Aid for Scientific Research for Young Scientist (A)( 25707031).

Figure captions

Fig. 1. (a) Schematic image of the crystal structure of LaOBiS$_{2-x}$Se$_x$. (b) Powder XRD patterns for the end members LaOBiS$_2$ and LaOBiSSe ($x$ = 0 and 1.0). (c) Se concentration dependence of lattice constant of $a$ for LaOBiS$_{2-x}$Se$_x$. (d) Se concentration dependence of lattice constant of $c$ for LaOBiS$_{2-x}$Se$_x$.

Fig. 2. (a) Temperature dependences of electrical resistivity for LaOBiS$_{2-x}$Se$_x$. (b) Temperature dependences of Seebeck coefficient for LaOBiS$_{2-x}$Se$_x$. (c) Temperature dependences of power factor for LaOBiS$_{2-x}$Se$_x$.

Fig. 3. (a) Temperature dependences of thermal conductivity for LaOBiS$_2$ and LaOBiSSe ($x$ = 0 and 1.0). (b) Temperature dependences of $ZT$ for LaOBiS$_{2-x}$Se$_x$.

Fig. 1(a)

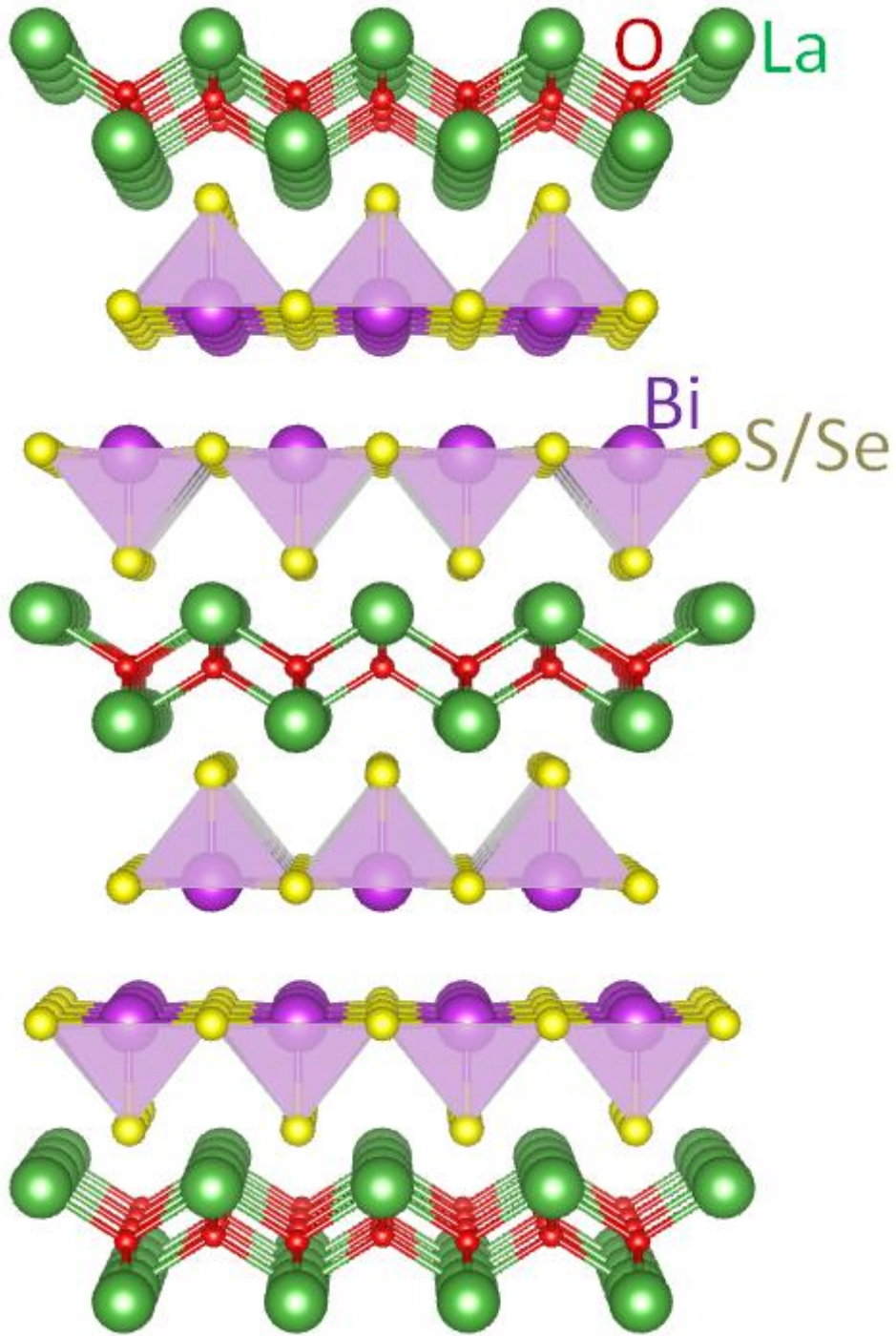

Fig. 1(b)

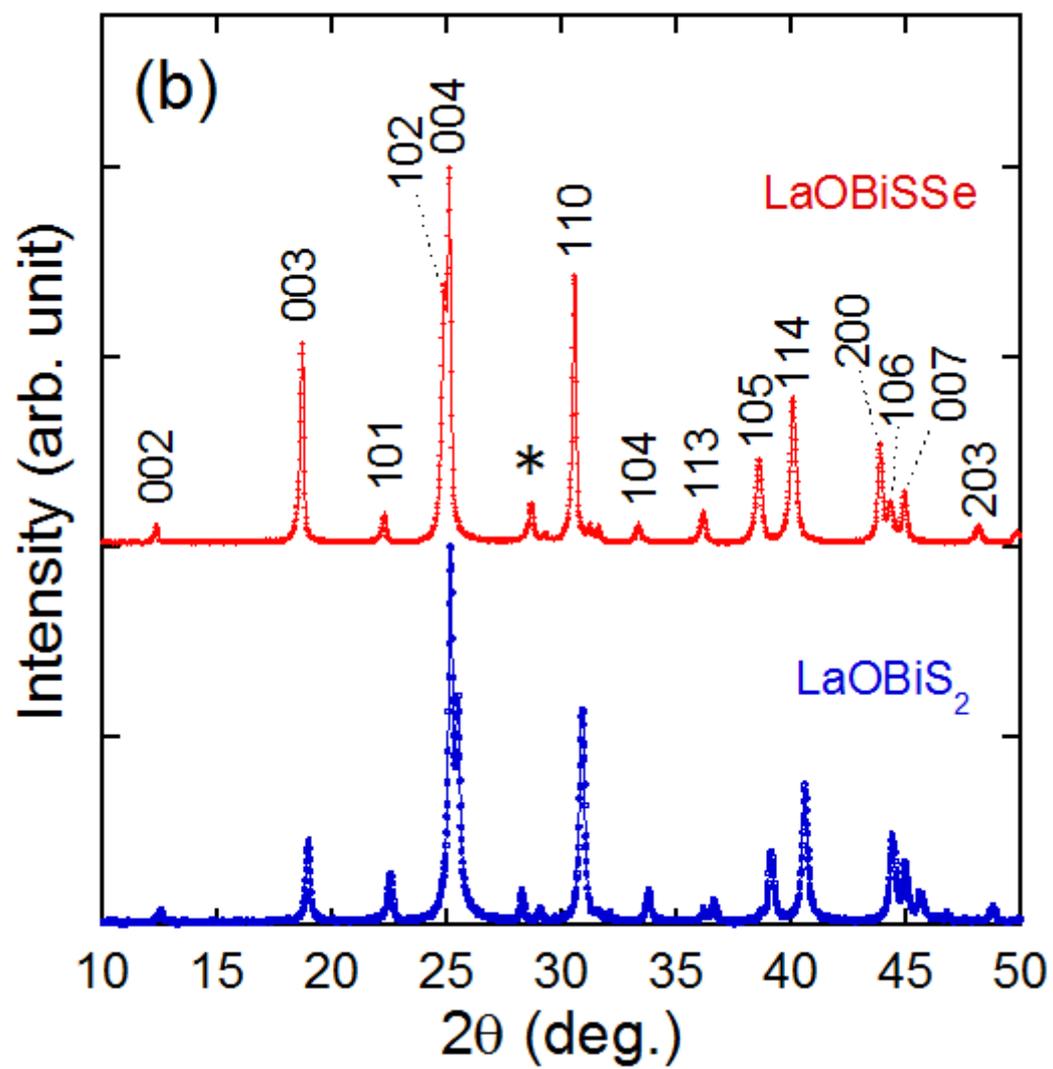

Fig. 1(c,d)

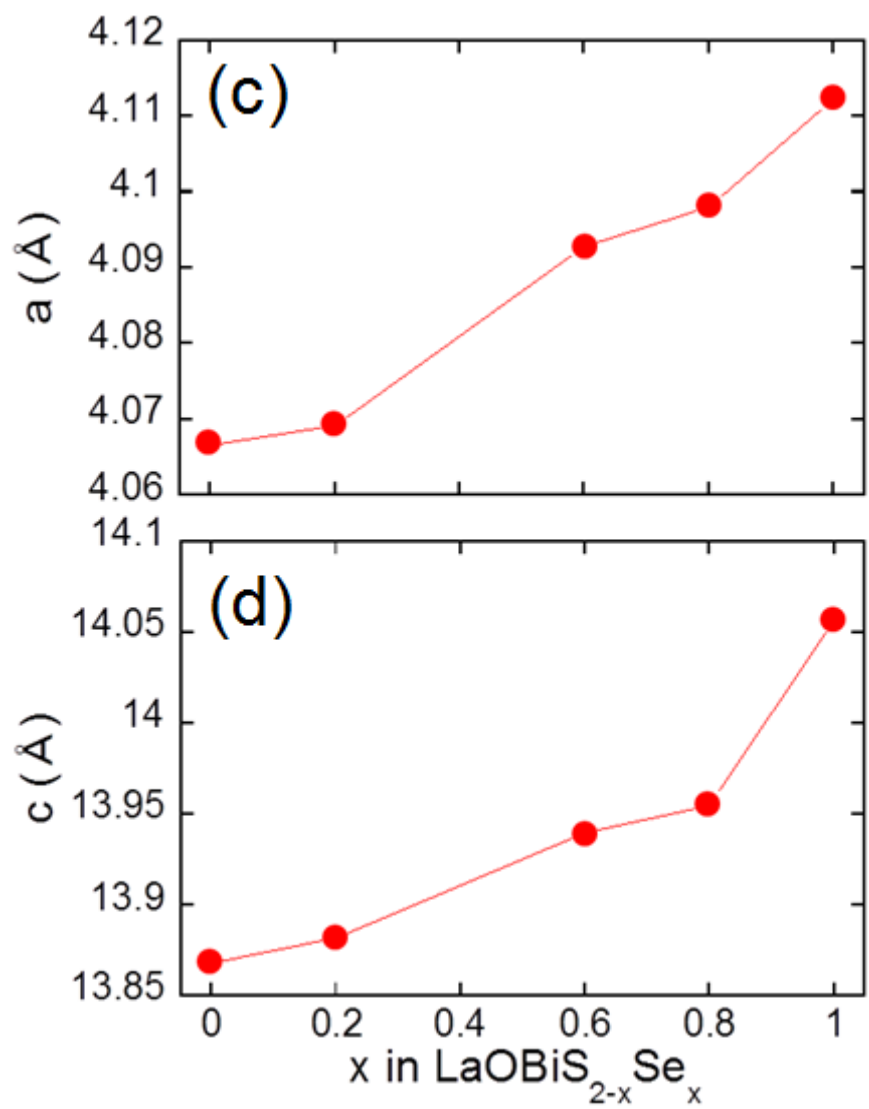

Fig. 4

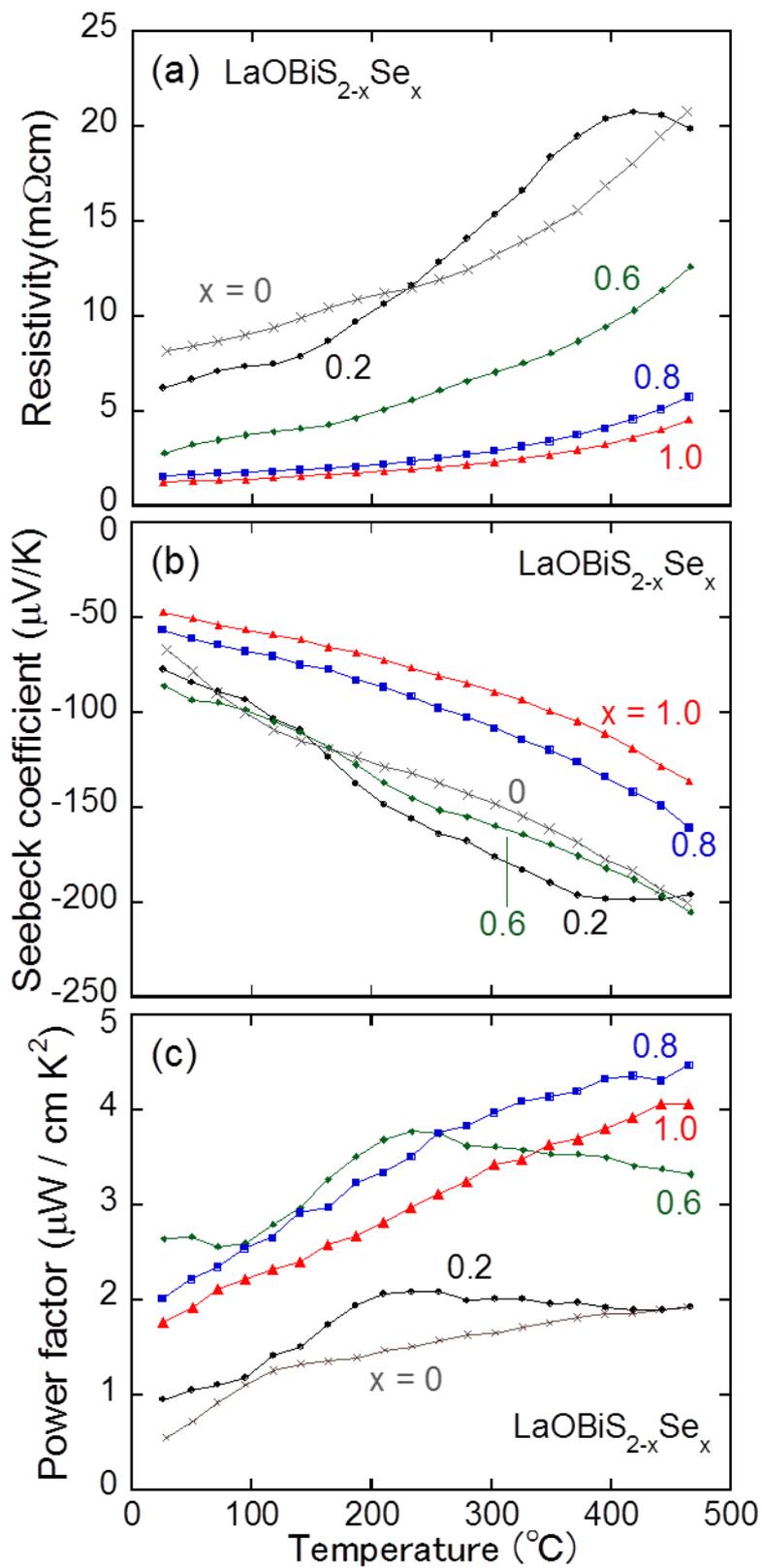

Fig. 5

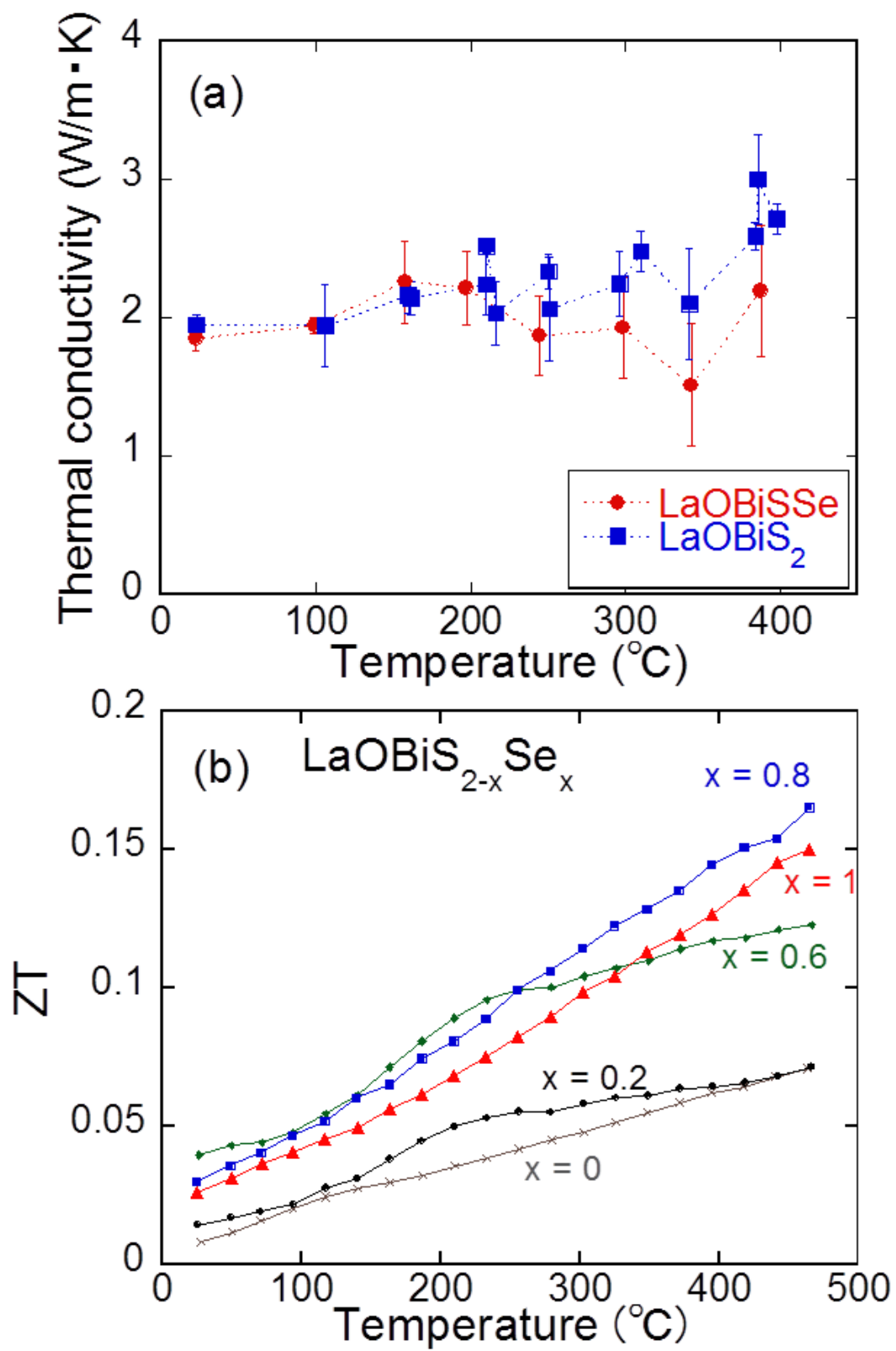